\documentclass[10pt,a4paper,onecolumn]{article}

\providecommand{\pandocbounded}[1]{#1}

\makeatletter

\let\olditem\item
\renewcommand{\item}[1][]{\olditem}
\makeatother

\usepackage{marginnote}
\usepackage{graphicx}
\usepackage{xcolor}
\usepackage{authblk,etoolbox}
\usepackage{titlesec}
\usepackage{calc}
\usepackage{tikz}
\usepackage{hyperref}
\hypersetup{colorlinks,breaklinks,
            urlcolor=[rgb]{0.0, 0.5, 1.0},
            linkcolor=[rgb]{0.0, 0.5, 1.0}}
\usepackage{caption}
\usepackage{tcolorbox}
\usepackage{amssymb,amsmath}
\usepackage{ifxetex,ifluatex}
\usepackage{seqsplit}
\usepackage{xstring}

\usepackage{float}
\let\origfigure\figure
\let\endorigfigure\endfigure
\renewenvironment{figure}[1][2] {
    \expandafter\origfigure\expandafter[H]
} {
    \endorigfigure
}

\usepackage{fixltx2e} % provides \textsubscript

\let\textttOrig=\texttt
\def\texttt#1{\expandafter\textttOrig{\seqsplit{#1}}}
\renewcommand{\seqinsert}{\ifmmode
  \allowbreak
  \else\penalty6000\hspace{0pt plus 0.02em}\fi}

\makeatletter
\let\href@Orig=\href
\def\href@Urllike#1#2{\href@Orig{#1}{\begingroup
    \def\Url@String{#2}\Url@FormatString
    \endgroup}}
\def\href@Notdoi#1#2{\def\tempa{#1}\def\tempb{#2}%
  \ifx\tempa\tempb\relax\href@Urllike{#1}{#2}\else
  \href@Orig{#1}{#2}\fi}
\def\href#1#2{%
  \IfBeginWith{#1}{https://doi.org}%
  {\href@Urllike{#1}{#2}}{\href@Notdoi{#1}{#2}}}
\makeatother

\newlength{\cslhangindent}
\newlength{\csllabelwidth}
\newenvironment{CSLReferences}[3] % #1 hanging-ident, #2 entry spacing
 {% don't indent paragraphs
  \setlength{\parindent}{0pt}
  \ifodd #1 \everypar{\setlength{\hangindent}{\cslhangindent}}\ignorespaces\fi
  \ifnum #2 > 0
  \setlength{\parskip}{#2\baselineskip}
  \fi
 }%
 {}
\usepackage{calc}

\usepackage[top=3.5cm, bottom=3cm, right=1.5cm, left=1.0cm,
            headheight=2.2cm, reversemp, includemp, marginparwidth=4.5cm]{geometry}

\titleformat{\section}
  {\normalfont\sffamily\Large\bfseries}
  {}{0pt}{}
\titleformat{\subsection}
  {\normalfont\sffamily\large\bfseries}
  {}{0pt}{}
\titleformat{\subsubsection}
  {\normalfont\sffamily\bfseries}
  {}{0pt}{}
\titleformat*{\paragraph}
  {\sffamily\normalsize}

\usepackage{fancyhdr}
\makeatletter
\let\ps@plain\ps@fancy
\fancyheadoffset[L]{4.5cm}
\fancyfootoffset[L]{4.5cm}

\definecolor{linky}{rgb}{0.0, 0.5, 1.0}

\newtcolorbox{repobox}
   {colback=red, colframe=red!75!black,
     boxrule=0.5pt, arc=2pt, left=6pt, right=6pt, top=3pt, bottom=3pt}

\newcommand{\ExternalLink}{%
   \tikz[x=1.2ex, y=1.2ex, baseline=-0.05ex]{%
       \begin{scope}[x=1ex, y=1ex]
           \clip (-0.1,-0.1)
               --++ (-0, 1.2)
               --++ (0.6, 0)
               --++ (0, -0.6)
               --++ (0.6, 0)
               --++ (0, -1);
           \path[draw,
               line width = 0.5,
               rounded corners=0.5]
               (0,0) rectangle (1,1);
       \end{scope}
       \path[draw, line width = 0.5] (0.5, 0.5)
           -- (1, 1);
       \path[draw, line width = 0.5] (0.6, 1)
           -- (1, 1) -- (1, 0.6);
       }
   }

\patchcmd{\@maketitle}{center}{flushleft}{}{}
\patchcmd{\@maketitle}{center}{flushleft}{}{}
\patchcmd{\@maketitle}{\LARGE}{\LARGE\sffamily}{}{}
\def\maketitle{{%
  
  \AB@maketitle}}
\makeatletter
\renewcommand\AB@affilsepx{ \protect\Affilfont}
\renewcommand\AB@affilnote[1]{{\bfseries #1}\hspace{3pt}}
\renewcommand{\affil}[2][]%
   {\newaffiltrue\let\AB@blk@and\AB@pand
      \if\relax#1\relax\def\AB@note{\AB@thenote}\else\def\AB@note{#1}%
        \setcounter{Maxaffil}{0}\fi
        \begingroup
        \let\href=\href@Orig
        \let\texttt=\textttOrig
        \let\protect\@unexpandable@protect
        \def\thanks{\protect\thanks}\def\footnote{\protect\footnote}%
        \@temptokena=\expandafter{\AB@authors}%
        {\def\\{\protect\\\protect\Affilfont}\xdef\AB@temp{#2}}%
         \xdef\AB@authors{\the\@temptokena\AB@las\AB@au@str
         \protect\\[\affilsep]\protect\Affilfont\AB@temp}%
         \gdef\AB@las{}\gdef\AB@au@str{}%
        {\def\\{, \ignorespaces}\xdef\AB@temp{#2}}%
        \@temptokena=\expandafter{\AB@affillist}%
        \xdef\AB@affillist{\the\@temptokena \AB@affilsep
          \AB@affilnote{\AB@note}\protect\Affilfont\AB@temp}%
      \endgroup
       \let\AB@affilsep\AB@affilsepx
}
\makeatother

\renewcommand\Affilfont{\sffamily\small\mdseries}
\ifnum 0\ifxetex 1\fi\ifluatex 1\fi=0 % if pdftex
  \usepackage[T1]{fontenc}
  \usepackage[utf8]{inputenc}

\else % if luatex or xelatex
  \ifxetex
    \usepackage{mathspec}
  \else
    \usepackage{fontspec}
  \fi
  \defaultfontfeatures{Ligatures=TeX,Scale=MatchLowercase}

\fi
\IfFileExists{upquote.sty}{\usepackage{upquote}}{}
\IfFileExists{microtype.sty}{%
\usepackage{microtype}
\UseMicrotypeSet[protrusion]{basicmath} % disable protrusion for tt fonts
}{}

\usepackage{hyperref}
\hypersetup{unicode=true,
            pdftitle={chatter: a Python library for applying information theory and AI/ML models to animal communication},
            pdfborder={0 0 0},
            breaklinks=true}
\let\addcontentslineOrig=\addcontentsline
\def\addcontentsline#1#2#3{\bgroup
  \let\texttt=\textttOrig\addcontentslineOrig{#1}{#2}{#3}\egroup}
\let\markbothOrig\markboth
\def\markboth#1#2{\bgroup
  \let\texttt=\textttOrig\markbothOrig{#1}{#2}\egroup}
\let\markrightOrig\markright
\def\markright#1{\bgroup
  \let\texttt=\textttOrig\markrightOrig{#1}\egroup}

\usepackage{graphicx,grffile}
\makeatletter
\def\maxwidth{\ifdim\Gin@nat@width>\linewidth\linewidth\else\Gin@nat@width\fi}
\def\maxheight{\ifdim\Gin@nat@height>\textheight\textheight\else\Gin@nat@height\fi}
\makeatother
\setkeys{Gin}{width=\maxwidth,height=\maxheight,keepaspectratio}
\IfFileExists{parskip.sty}{%
\usepackage{parskip}
}{% else
\setlength{\parindent}{0pt}
\setlength{\parskip}{6pt plus 2pt minus 1pt}
}
\ifx\paragraph\undefined\else
\let\oldparagraph\paragraph
\renewcommand{\paragraph}[1]{\oldparagraph{#1}\mbox{}}
\fi
\ifx\subparagraph\undefined\else
\let\oldsubparagraph\subparagraph
\renewcommand{\subparagraph}[1]{\oldsubparagraph{#1}\mbox{}}
\fi

\title{\texttt{chatter}: a Python library for applying information
theory and AI/ML models to animal communication}

        \author{Mason Youngblood}
    
      \affil[]{Institute for Advanced Computational Science, Stony Brook
University, USA masonyoungblood@gmail.com}
  \date{\vspace{-5ex}}

\begin{document}
\maketitle

\marginpar{
  \sffamily\small

  \vspace{2mm}

  {\bfseries Software}
  \begin{itemize}
    \setlength\itemsep{0em}
    \item \href{https://github.com/masonyoungblood/chatter}{\color{linky}{GitHub}} \ExternalLink
    \item \href{https://masonyoungblood.github.io/chatter}{\color{linky}{Documentation}} \ExternalLink
    \item \href{https://pypi.org/project/chatter-pkg/}{\color{linky}{PyPI}} \ExternalLink
  \end{itemize}

  \vspace{2mm}

  {\bfseries Date:} 10 December 2025

}

\section{Summary}\label{summary}

The study of animal communication often involves categorizing units into
types (e.g.~syllables in songbirds, or notes in humpback whales). While
this approach is useful in many cases, it necessarily flattens the
complexity and nuance present in real communication systems.
\texttt{chatter} is a new Python library for analyzing animal
communication in continuous latent space using information theory and
modern machine learning techniques. It is taxonomically agnostic, and
has been tested with the vocalizations of birds, bats, whales, and
primates. By leveraging a variety of different architectures, including
variational autoencoders and vision transformers, \texttt{chatter}
represents vocal sequences as trajectories in high-dimensional latent
space, bypassing the need for manual or automatic categorization of
units. The library provides an end-to-end workflow---from preprocessing
and segmentation to model training and feature extraction---that enables
researchers to quantify the complexity, predictability, similarity, and
novelty of vocal sequences.

\section{Statement of Need}\label{statement-of-need}

In recent years, animal behaviorists have started to use machine
learning to project vocalizations into high-dimensional latent space
(Goffinet et al., 2021; Merino Recalde, 2023; Sainburg et al., 2020).
However, these methods are typically used to identify categories of
vocalizations instead of enabling analysis of vocal sequences as
continuous vectors, with some exceptions (Alam et al., 2024).
\texttt{chatter} addresses this gap by providing an accessible, modular,
and easy-to-use interface for applying these techniques. It complements
existing tools like \texttt{pykanto} (Merino Recalde, 2023) and
\texttt{AVGN} (Sainburg et al., 2020), which focus on discrete analysis,
by offering a parallel workflow for continuous analysis. In doing so, it
lowers the barrier to entry for researchers who are interested in
computational methods but may not have the time or expertise to assemble
a machine learning pipeline from scratch. The library is designed to be
flexible enough for advanced users while providing sensible defaults for
novices.

The core workflow of \texttt{chatter} has three steps, all managed by a
configuration dictionary. First, users initialize the \texttt{Analyzer}
class, which takes parameters like minimum and maximum frequency,
amplitude thresholds, and denoising settings. The \texttt{Analyzer}
handles the raw audio, preprocessing it with noise reduction, filtering,
and compression, and segmenting it into individual acoustic units based
on amplitude or other criteria. It also gives users working with
birdsong data the ability to identify their focal species using BirdNET
(Kahl et al., 2021). The resulting units are converted into
spectrograms.

\begin{figure}
\centering
\pandocbounded{\includegraphics[keepaspectratio,alt={A basic diagram of the chatter workflow, showing the progression from spectrograms to latent features to visualizations in 2D space. Note that all of the information theoretic analysis occurs in original latent space, not in the reduced 2D space.}]{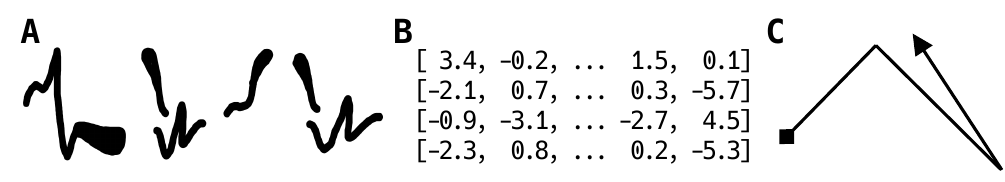}}
\caption{A basic diagram of the \protect\texttt{chatter} workflow,
showing the progression from spectrograms to latent features to
visualizations in 2D space. Note that all of the information theoretic
analysis occurs in original latent space, not in the reduced 2D space.}
\end{figure}

Next, the \texttt{Trainer} class is used to train a convolutional
variational autoencoder to learn a high-dimensional latent
representation of the vocalizations (Goffinet et al., 2021). This
unsupervised approach is able to describe complex vocalizations in a
relatively lossless way---allowing users to reconstruct the original
sound from a set of learned acoustic features. These latent features are
then extracted for all acoustic units. Users also have the option of
collecting latent features by applying the DINOv3 vision transformer to
spectrograms, which works especially well when treating sequences as
units (i.e.~working with spectrograms of whole sequences rather than
units within sequences) (Siméoni et al., 2025). Finally, the library
provides tools to visualize this latent space using dimensionality
reduction techniques such as PaCMAP (Wang et al., 2021), an extension of
UMAP that better captures global structure. Figure 2 illustrates this
output for Cassin's vireo syllables (recordings from Hedley (2016)).

\begin{figure}
\centering
\pandocbounded{\includegraphics[keepaspectratio,alt={The latent space of Cassin's vireo syllables. The plot visualizes the syllables in a 2D latent space produced by applying PaCMAP to the latent features from a variational autoencoder, with representative spectrograms overlaid.}]{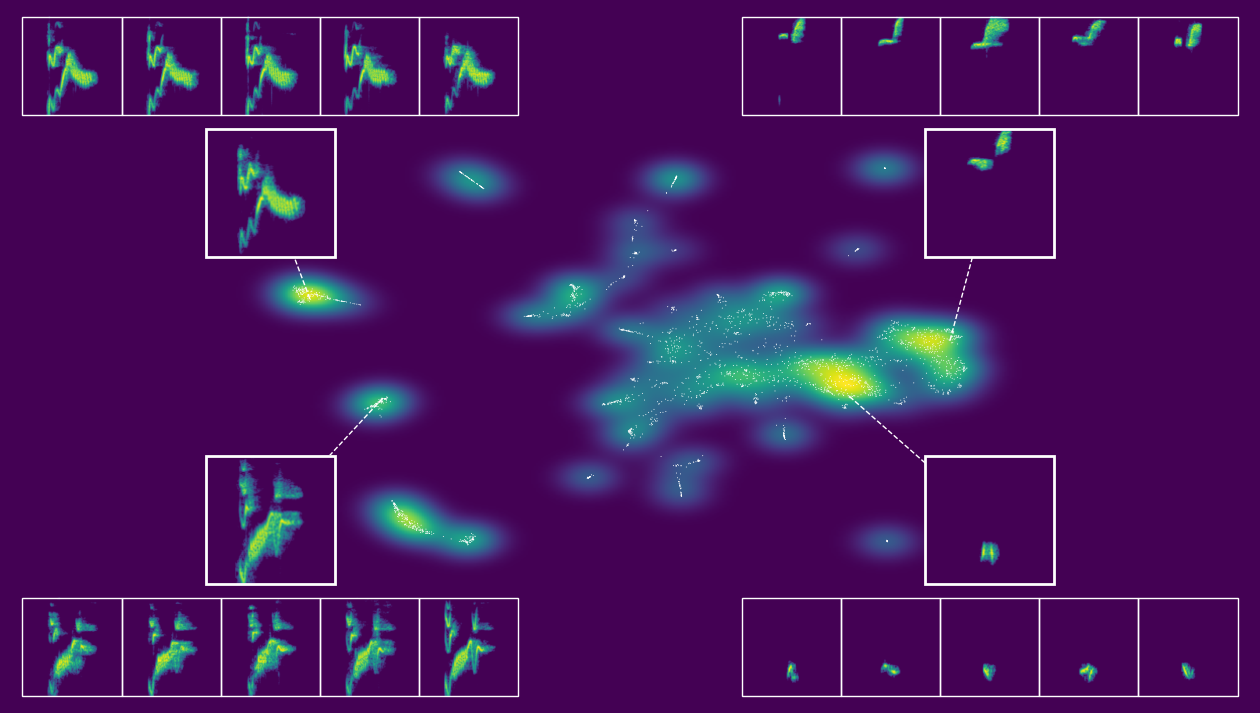}}
\caption{The latent space of Cassin's vireo syllables. The plot
visualizes the syllables in a 2D latent space produced by applying
PaCMAP to the latent features from a variational autoencoder, with
representative spectrograms overlaid.}
\end{figure}

Finally, the \texttt{FeatureProcessor} class provides a variety of
functions for downstream analysis. Complexity is quantified by
calculating the path length (cosine distance) between consecutive
acoustic units in latent space, normalized by duration, offering a
continuous measure of acoustic diversity over time (Alam et al., 2024)
similar to previous methods based on spectrogram cross-correlation
(Sawant et al., 2022). To assess predictability, \texttt{chatter} trains
a vector autoregression model on sequences of latent features (Sims,
1980). This model learns temporal dependencies and quantifies the
probability of future states. \texttt{chatter} also estimates the
commonness and rarity of acoustic units using \texttt{denmarf}, a
density estimation library that uses masked autoregressive flows to
compute density in high-dimensional space (Papamakarios et al., 2017).
This provides a measure of how common or rare a given vocalization is
relative to the learned distribution. Finally, similarity between whole
sequences of units is computed by applying dynamic time warping to the
sequences of latent vectors (i.e.~panel B in Figure 1). Collectively,
these measures provide a fairly complex and nuanced description of
animal vocal sequences.

\section*{References}\label{references}
\addcontentsline{toc}{section}{References}

\protect\phantomsection\label{refs}
\begin{CSLReferences}{1}{0}
\par\noindent Alam, D., Zia, F., \& Roberts, T. F. (2024). The hidden fitness of the
male zebra finch courtship song. \emph{Nature}, \emph{627}, 851--857.
\url{https://doi.org/10.1038/s41586-024-07207-4}

\par\noindent Goffinet, J., Brudner, S., Mooney, R., \& Pearson, J. (2021).
Low-dimensional learned feature spaces quantify individual and group
differences in vocal repertoires. \emph{eLife}, \emph{10}, e67855.
\url{https://doi.org/10.7554/eLife.67855}

\par\noindent Hedley, R. W. (2016). Complexity, predictability, and time homogeneity
of syntax in the songs of cassin's vireo. \emph{PLOS ONE}, \emph{11}(3),
e0150822. \url{https://doi.org/10.1371/journal.pone.0150822}

\par\noindent Kahl, S., Wood, C. M., Eibl, M., \& Klinck, H. (2021). BirdNET: A deep
learning solution for avian diversity monitoring. \emph{Ecological
Informatics}, \emph{61}, 101236.
\url{https://doi.org/10.1016/j.ecoinf.2021.101236}

\par\noindent Merino Recalde, N. (2023). Pykanto: A python library to accelerate
research on wild bird song. \emph{Methods in Ecology and Evolution},
\emph{14}(6), 1639--1648. \url{https://doi.org/10.1111/2041-210X.14155}

\par\noindent Papamakarios, G., Pavlakou, T., \& Murray, I. (2017). Masked
autoregressive flow for density estimation. In I. Guyon, U. V. Luxburg,
S. Bengio, H. Wallach, R. Fergus, S. Vishwanathan, \& R. Garnett (Eds.),
\emph{Advances in neural information processing systems} (Vol. 30).
\url{https://proceedings.neurips.cc/paper_files/paper/2017/file/6c1da886822c67822bcf3679d04369fa-Paper.pdf}

\par\noindent Sainburg, T., Thielk, M., \& Gentner, T. Q. (2020). Finding,
visualizing, and quantifying latent structure across diverse animal
vocal repertoires. \emph{PLOS Computational Biology}, \emph{16}(10),
e1008228. \url{https://doi.org/10.1371/journal.pcbi.1008228}

\par\noindent Sawant, S., Arvind, C., Joshi, V., \& Robin, V. V. (2022). Spectrogram
cross-correlation can be used to measure the complexity of bird
vocalizations. \emph{Methods in Ecology and Evolution}, \emph{13}(2),
459--472. \url{https://doi.org/10.1111/2041-210X.13765}

\par\noindent Siméoni, O., Vo, H. V., Seitzer, M., Baldassarre, F., Oquab, M., Jose,
C., Khalidov, V., Szafraniec, M., Yi, S., Ramamonjisoa, M., Massa, F.,
Haziza, D., Wehrstedt, L., Wang, J., Darcet, T., Moutakanni, T.,
Sentana, L., Roberts, C., Vedaldi, A., \ldots{} Bojanowski, P. (2025).
\emph{DINOv3}. \url{https://arxiv.org/abs/2508.10104}

\par\noindent Sims, C. A. (1980). Macroeconomics and reality. \emph{Econometrica},
\emph{48}(1), 1--48. \url{https://doi.org/10.2307/1912017}

\par\noindent Wang, Y., Huang, H., Rudin, C., \& Shaposhnik, Y. (2021). Understanding
how dimension reduction tools work: An empirical approach to ceciphering
t-SNE, UMAP, TriMap, and PaCMAP for data visualization. \emph{Journal of
Machine Learning Research}, \emph{22}(201), 1--73.
\url{http://jmlr.org/papers/v22/20-1061.html}

\end{CSLReferences}

\end{document}